\documentclass[12pt,a4paper]{article}
\usepackage{graphicx}
\usepackage{axodraw}
\usepackage{amsmath}
\usepackage{amssymb}
\newcommand{\tr}{{\rm Tr}}
\newcommand{\gev}{\mbox{{\rm GeV}}}
\newcommand{\tev}{\mbox{{\rm TeV}}}
\newcommand{\hc}{\mbox{{\rm h.\,c.}}}
\newcommand{\abs}[1]{\left\vert#1\right\vert}

\newcommand{\sla}[1]{#1 \!\!\!\! \diagup}
\begin{document}
\begin{titlepage}

\begin{flushright}
\end{flushright}

\vskip 1.5cm

\begin{center}
{\Large \bf  Breaking the electroweak symmetry and supersymmetry
by a compact extra dimension}

\vskip 1.0cm

{\bf Riccardo Barbieri, Guido Marandella, Michele Papucci }

\vskip 0.5cm

{\it Scuola Normale Superiore and INFN, \\ Piazza dei Cavalieri 7,
                 I-56126 Pisa, Italy}\\

\vskip 1.0cm

\begin{abstract}
We revisit in some more detail a recent specific proposal for the
breaking of the electroweak symmetry and of supersymmetry by a
compact extra dimension. Possible mass terms for the Higgs and the
matter hypermultiplets are considered and their effects on the
spectrum analyzed. Previous conclusions are reinforced and put on
firmer ground.
\end{abstract}

\end{center}
\end{titlepage}

\section{Introduction and motivation}

The ElectroWeak Symmetry Breaking (EWSB) remains an unsettled
central problem in particle physics. No clear experimental signal
has emerged yet which point towards a specific physical
description of EWSB. The Higgs boson has not been found, nor any
supersymmetric particle, which, as believed by many, could play a
crucial role in triggering EWSB. Similarly any possible mechanism
of dynamical EWSB, if realized, is at least well hidden in the
relevant data so far. This is where we stand at the moment. It is
very likely, on the other hand, that the progression of the
upgraded Tevatron runs and, especially, the coming in operation of
LHC will change the situation in this decade, making available
crucial data on the physics of EWSB. All this motivates further
thoughts on this problem.

Where to look for an orientation, however? Theoretically, the
quadratic divergence of the Higgs mass in the Standard Model (SM)
remains a crucial aspect of EWSB. This is neither new, however,
nor sufficiently discriminating among alternative theoretical
ideas. More significant, maybe, is the impressive series of
ElectroWeak Precision Tests (EWPT) performed in the nineties
mostly at LEP but also at the Tevatron and at SLC. As well known,
these data brilliantly confirm the SM at the level of the pure
electroweak radiative corrections. Although indirectly, this
suggests that any drastic departure from the SM cannot occur, if
it does at all, below a few {\tev}. Furthermore, always
indirectly, an evidence emerges from the same data in favor of a
light Higgs in the hundred {\gev} range. Here we stick to this
interpretations of the EWPT, not unavoidable but plausible: there
is indeed a light Higgs, close to the direct lower bound of about
110 {\gev}, while physics remains perturbative up to 2-3 {\tev}
energies at least.

This view would have a problem, however, if we also supposed, at
the same time, that no new weakly interacting particle were present
below the cut off scale $\Lambda$, at or above 2-3 {\tev}. The
dominant radiative correction to the Higgs mass, from the top
loop, cut-off at $\Lambda$ would in fact be ($G_F$ is the Fermi
constant and $m_t$ the top mass):
\begin{equation}
\delta m_H^2 (top) = \frac{3}{\sqrt 2 \pi^2} G_F \, m_t^2 \,
\Lambda^2 = \left( 0.9 \, {\tev} \right) ^2 \left(\frac{\Lambda}{3
\, {\tev}} \right)^2  \label{eq:corr-mh-top}
\end{equation}
at least about $100$ times bigger than the supposed physical Higgs
squared mass. The usual hierarchy problem, coupled with the
knowledge of the top mass, has acquired a ``low energy'' aspect.
We underline that this was not the case, a decade ago, when the
EWPT were not available and the top mass was not known. Given the
special role of LEP in the EWPT, one can call this the ``LEP
paradox'' \cite{Barbieri:2000gf}.

All this sounds pretty familiar as an argument in favor of the
existence of superpartners, and maybe it is. With the introduction
of the loops of the stop, of mass $m_{\tilde t}$,
(\ref{eq:corr-mh-top}) turns into:
\begin{equation}
\delta m_H^2 (top\, - \, stop) = \frac{3}{\sqrt 2 \pi^2} G_F \,
m_t^2 \, m_{\tilde t}^2 \, \log \frac{\Lambda^2}{m_{\tilde t}^2} 
= 0.1 \, m_{\tilde t}^2 \, \log \frac{\Lambda^2}{m_{\tilde t}^2}
\label{eq:corr-mh-top/stop}
\end{equation}
still divergent, but only logarithmically. The stop and the other
superpartners might exist then, but where? It is in the very
spirit of this entire argument that the correction
(\ref{eq:corr-mh-top/stop}) and the other contributions to $m_H^2$
should not exceed significantly the $(100 {\gev})^2$ range without
an accidental tuning among the different parameters involved. In
turn, by inspection of definite supersymmetric extensions of the
SM, this has led to the expectation that some superpartner,as the
Higgs itself, had to
be discovered, in particular, at LEP, which did not happen. Since
this was an expectation and not a theorem, the failure to find
supersymmetry at LEP is not of immediate interpretation either.
With some amount of tuning in the space of parameters, standard
superpartners, with masses close to the current lower bounds, can
certainly exist, ready to be found at the upgraded Tevatron and/or
at LHC. The success of gauge coupling unification supports this
view. On the bad side, however, by allowing an increasing amount of tuning,
all superpartners could escape detection even at LHC. 

All this justifies, in our opinion, the exploration of alternative
possibilities: we seek a model with a naturally light Higgs, in
the 100 {\gev} region, perturbative up to a few {\tev} at least
and with a structure in between possibly determined in terms of a
minimum number of parameters. This has motivated the
proposal of Ref. \cite{Barbieri:2001vh}. In this paper we return
to it in some more detail, also in view of the contents of Refs.
\cite{Ghilencea:2001bw,Barbieri:2001cz,Scrucca:2002eb,Barbieri:2002ic}.

The structure of the paper is the following. In section
\ref{sec:LEP-paradox} we recall the solution of the LEP paradox
proposed in Ref. \cite{Barbieri:2001vh}. Section \ref{sec:model}
contains a description of the complete extension of the SM and of
the in principle relevant parameter space. This includes suitable
mass terms for the matter and Higgs hypermultiplets as discussed
in \cite{Barbieri:2002ic}. In section \ref{sec:EWSB} we give a
detailed discussion of EWSB with the inclusion of mass terms, small
relative to $1/R$. The spectrum of the model and the consequent
phenomenological implications are summarized in section
\ref{sec:phenomenology}.

\section{A solution of the LEP paradox} \label{sec:LEP-paradox}

\begin{figure}[b]
\begin{center}
\begin{picture}(160,140)(0,20)
  \LongArrow(80,130)(30,130)
  \LongArrow(80,130)(130,130)
  \Text(15,125)[b]{$f$}
  \Text(80,135)[b]{SUSY}
  \Text(145,125)[b]{$\tilde f$}
  \LongArrow(80,30)(30,30)
  \LongArrow(80,30)(130,30)
  \Text(15,25)[b]{$f^c$}
  \Text(80,35)[b]{SUSY}
  \Text(145,25)[b]{$\tilde f^c$}
  \LongArrow(15,80)(15,120)
  \LongArrow(15,80)(15,40)
  \Text(5,80)[b]{5D}
  \LongArrow(145,80)(145,120)
  \LongArrow(145,80)(145,40)
  \Text(155,80)[b]{5D}
\end{picture}
\caption{Component diagram of the hypermultiplet in 5D.}
\label{fig:5d-hyper-scheme}
\end{center}
\end{figure}
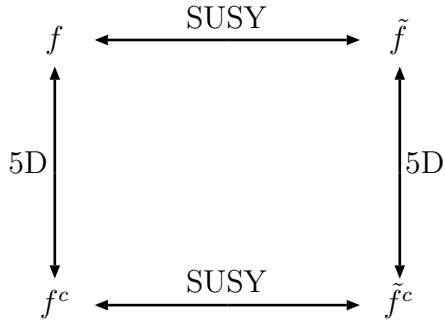

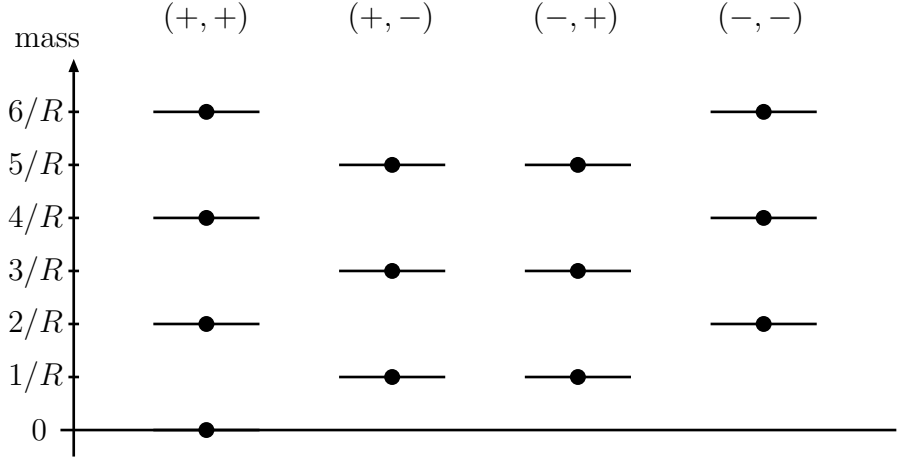
\begin{figure}[t]
\begin{center}
\begin{picture}(350,190)(-10,-25)
  \Line(5,0)(320,0)
  \LongArrow(10,-10)(10,140)
  \Text(0,145)[b]{mass}
  \Text(0,0)[r]{$0$}
  \Line(8,20)(12,20)    \Text(6,20)[r]{$1/R$}
  \Line(8,40)(12,40)    \Text(6,40)[r]{$2/R$}
  \Line(8,60)(12,60)    \Text(6,60)[r]{$3/R$}
  \Line(8,80)(12,80)    \Text(6,80)[r]{$4/R$}
  \Line(8,100)(12,100)  \Text(6,100)[r]{$5/R$}
  \Line(8,120)(12,120)  \Text(6,120)[r]{$6/R$}
  \Text(60,150)[b]{$(+,+)$}
  \Line(40,0)(80,0)      \Vertex(60,0){3}
  \Line(40,40)(80,40)    \Vertex(60,40){3}
  \Line(40,80)(80,80)    \Vertex(60,80){3}
  \Line(40,120)(80,120)  \Vertex(60,120){3}
  \Text(130,150)[b]{$(+,-)$}
  \Line(110,20)(150,20)    \Vertex(130,20){3}
  \Line(110,60)(150,60)    \Vertex(130,60){3}
  \Line(110,100)(150,100)  \Vertex(130,100){3}
  \Text(200,150)[b]{$(-,+)$}
  \Line(180,20)(220,20)    \Vertex(200,20){3}
  \Line(180,60)(220,60)    \Vertex(200,60){3}
  \Line(180,100)(220,100)  \Vertex(200,100){3}
  \Text(270,150)[b]{$(-,-)$}
  \Line(250,40)(290,40)    \Vertex(270,40){3}
  \Line(250,80)(290,80)    \Vertex(270,80){3}
  \Line(250,120)(290,120)  \Vertex(270,120){3}
\end{picture}
\caption{Tree-level KK mass spectrum of a multiplet (vector,
matter or Higgs) with the indicated boundary conditions.}
\label{fig:spectrum}
\end{center}
\end{figure}

We suppose that supersymmetry is relevant to solve the LEP
paradox, as defined above, and that the left handed top, with its
doublet $Q = (t,b)_L$, and the right handed top $t_R$ live in 5
dimensions of coordinates $(x_\mu,y)$. For every matter Weyl
spinor $f$ this amounts to introducing a hypermultiplet of
$(x_\mu,y)$-dependent fields according to the scheme of Fig.
\ref{fig:5d-hyper-scheme}, where $f^c$ denotes a spinor with the
same chirality of $f$ but opposite quantum numbers.
The 5th dimension is viewed as a segment of length $\pi R/2$ with
Dirichlet $(+)$ or Neumann $(-)$ conditions on the two boundaries:
$(+,+)$, $(-,-)$, $(+,-)$, $(-,+)$ respectively for the fields
$f$, $f^c$, $\tilde f$, $\tilde f^c$. This is the unique way,
consistently with the symmetries of the free 5D Lagrangian, to
obtain a spectrum with a single massless mode for the fermion $f$
only. The spectrum for the entire hypermultiplet with these
boundary conditions is given in Fig. \ref{fig:spectrum}. All
fields are periodic over a circle of radius R. The 5th dimension
can be viewed as  compactified on a $S^1 /(Z_2 \times Z_2')$
orbifold where $Z_2$ and $Z_2'$ are the reflections around the two
boundaries at $y=0$ and $y=\pi R/2$. Supersymmetry is broken {\`a} la
Scherk-Schwarz by the boundary conditions.

Consistently with supersymmetry, the top Yukawa coupling can be
introduced as a superpotential term localized at one of the
boundaries, say $y=0$,
\begin{equation}
\mathcal{L}_Y = \int \textrm d y \, \delta(y) \int \textrm d ^2
\theta \, \lambda_t \, \hat h \, \hat Q \, \hat U + \hc
\label{eq:yukawa-term}
\end{equation}
where $\hat h$, $\hat Q$, $\hat U$ are $N=1$ chiral multiplets. In
particular $\hat Q$ and $\hat U$ each contain the fields $f$ and
$\tilde f$ of Fig. \ref{fig:5d-hyper-scheme} with the
corresponding quantum numbers. It is irrelevant at this stage
whether $\hat h$ does or does not have a y-dependence. We assume
that the scalar $\hat h$ contains a y-independent component
$h^0(x)$ which plays the role of the standard Higgs field.

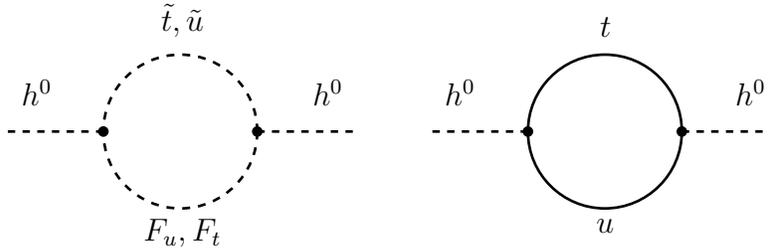
\begin{figure}[htb]
\begin{center}
\begin{picture}(270,110)(-265,-45)
  \DashLine(-265,10)(-229,10){3} \Vertex(-229,10){2}
  \Text(-255,20)[b]{$h^0$}
  \DashCArc(-200,10)(29,0,360){3}
  \Text(-200,46)[b]{$\tilde t, \tilde u$}
  \Text(-200,-23)[t]{$F_u, F_t$}
  \DashLine(-171,10)(-135,10){3} \Vertex(-171,10){2}
  \Text(-145,20)[b]{$h^0$}
  \DashLine(-105,10)(-69,10){3} \Vertex(-69,10){2}
  \Text(-95,20)[b]{$h^0$}
  \CArc(-40,10)(29,0,360)
  \Text(-40,46)[b]{$t$}
  \Text(-40,-23)[t]{$u$}
  \DashLine(-11,10)(25,10){3} \Vertex(-11,10){2}
  \Text(15,20)[b]{$h^0$}
\end{picture}
\caption{One-loop diagrams contributing to the mass squared of the
Higgs boson.} \label{fig:1loop-diagrams}
\end{center}
\end{figure}

We are in the position to compute the one loop contribution to the
Higgs mass due to the coupling (\ref{eq:yukawa-term}). This is
most readily done by means of the propagators in mixed $(p_\mu,y)$
space $G_i (p;y,y')$ for the different components of the
superfield, $i=f,\tilde f,F$ \cite{Arkani-Hamed:2001mi}. Corresponding to the diagrams of
Fig. \ref{fig:1loop-diagrams} one has:
\begin{equation}
\delta m_H^2 = 3 \lambda_t^2 \int \frac{{\rm d}^4 p}{(2 \, \pi)^4}
\left[ - \tr (G_t(p) \, G_u(p)) + G_{F_u} (p) \, G_{\tilde t} (p)
+ G_{F_t} (p) \, G_{\tilde u} (p) \right] \label{eq:corr-m2}
\end{equation}
where $G_i (p) = G_i (p;0,0)$. Using (\ref{eq:prop-M0-v0}) of
Appendix \ref{sec:app-propag} we obtain \cite{Barbieri:2001vh}
\begin{eqnarray}
\delta m_H^2 &=& - \frac{3 \, \widehat{y}_t^2}{16 \, R^2} \int_0^\infty {\rm
d}x \, x^3 \left[ \coth^2 \left(\frac{\pi x}{2}\right) - \tanh^2
\left(\frac{\pi x}{2}\right) \right] \nonumber \\
&=& - \frac{63 \, \zeta (3)}{8 \, \pi^4} \frac{\widehat{y}_t^2}{R^2}
\label{eq:corr-m2-result}
\end{eqnarray}
where $\zeta(3)=1.20$ and $\widehat{y}_t = \lambda_t / (2 \pi
R)^{3/2}$ is the top Yukawa coupling in 4D (anticipating a
y--dependent Higgs field as well). The finiteness of
(\ref{eq:corr-m2-result}) is a consequence of local supersymmetry
conservation in 5D, as discussed below.

\subsection{The relation between the compactification scale and
the cut-off}\label{sec:rel-cutoff}

The finiteness of (\ref{eq:corr-m2-result}) and the spectrum in
Fig. \ref{fig:spectrum}, with all extra particles in the top
hypermultiplet living at or above the compactification scale
$1/R$, look as a right step in the direction of solving the LEP
paradox. The price to be paid, however, is the
non-renormalizability of the coupling (\ref{eq:yukawa-term}) in
5D. Any model that incorporates the physics of section
\ref{sec:LEP-paradox} must be thought of as an effective field
theory valid up to some cut-off scale $\Lambda$. This is not
necessarily a problem, however, as long as $\Lambda$ is itself not
lower than \mbox{2-3 {\tev}} and is sufficiently bigger than the
compactification scale $1/R$ so that equation
(\ref{eq:corr-m2-result}), or similar ones, remain quantitatively
meaningful in the usual sense of effective field theories.

The relation between $1/R$ and $\Lambda$ can be fixed by requiring
that the top Yukawa coupling in (\ref{eq:yukawa-term}) becomes non
perturbative at $\Lambda$, taking into account the increasing
number of states whose thresholds are crossed at every unit of
$1/R$. With this assumption, the value of $\widehat y_t$ at
$\Lambda$ can either be estimated by means of usual dimensional
arguments, properly adapted to 5D \cite{Chacko:1999hg},
\begin{equation}
\widehat y_t(\Lambda) \simeq \frac{1}{16 \, \pi^2} \left(\frac{24 \,
\pi^3}{2 \, \pi \, \Lambda R} \right)^{3/2} \simeq 8.2 \, (\Lambda
R)^{-3/2} \label{eq:rel5D-yt-r-lambda}
\end{equation}
or by noticing that the expansion parameter in a 4D calculation
involving the top Yukawa coupling is\footnote{A factor of 2 is
included to account for the coupling of the non-zero KK modes,
$\sqrt 2$ times stronger than for the zero mode.}
\begin{equation}
\frac{2 \widehat y_t^2}{16 \, \pi^2} ( \, N_{KK})^3 \nonumber
\end{equation}
where $N_{KK} \simeq \Lambda R$ is the number of modes below
$\Lambda$, hence
\begin{equation}
\widehat y_t(\Lambda) \simeq \frac{4 \, \pi}{\sqrt 2}
\left(\frac{1}{ \Lambda R} \right)^{3/2} \simeq 8.9 \, (\Lambda
R)^{-3/2} \label{eq:rel4D-yt-r-lambda}
\end{equation}
Matching this value with the measured top Yukawa coupling at the
weak scale gives $\Lambda R \simeq 5$ \cite{Barbieri:2001vh}.
Note that $\widehat y_t$ at $\Lambda$ has not increased from 1 by more than
$20\%$ or so. The one-loop evolved $\widehat y_t$ starts growing rapidly at
$\Lambda \simeq 6/R$. From the 4D viewpoint, it is the
multiplicity of states, rather than the increase of $\widehat y_t$ itself,
that causes the loss of perturbativity.

Is $\Lambda R \simeq 5$ enough to defend the predictivity of an
equation like (\ref{eq:corr-m2-result})? We claim that it is, as
it can be checked by writing the most general Lagrangian,
involving the top and the Higgs fields, consistent with the
various symmetries and containing operators of arbitrarily high
dimensions, all assumed to saturate perturbation theory at
$\Lambda$. The corrections that these extra couplings induce are
not large. The value of $\Lambda$ itself, or of $1/R$, will be set
in the following. We also anticipate that the gauge couplings,
growing more slowly than $\widehat y_t$, remain perturbative below or at
$\Lambda$.

\section{The complete model and the relevant parameter space}
\label{sec:model}

The most straightforward way to include the gauge and Higgs
multiplets is to take every field in 5D. With matter and gauge
fields in 5D, (discrete) momentum conservation holds in the 5th
direction, thus weakening the lower bounds on $1/R$. Furthermore,
it is essential that the Higgs too lives in 5D if one does not
want to duplicate the Higgs multiplets as in standard
supersymmetric models (see below).

The parity assignments of the fields in the gauge multiplet, a 4D
vector $A_ \mu$, a 4D complex scalar $\varphi = \frac{1}{\sqrt 2}
(\Sigma + i A_5)$ and two Weyl spinors $\tilde \lambda$, $\tilde
\lambda ^c$ for every generator of the gauge group $SU(3) \times
SU(2) \times U(1)$ are fixed to be $(+,+)$, $(-,-)$, $(+,-)$ and
$(-,+)$ respectively by Lorenz, gauge and supersymmetry invariance
in 5D. Hence, from Fig. \ref{fig:spectrum}, the masslessness of
the vector zero modes only.

As to the Higgs hypermultiplet two choices are in principle
possible for the parity assignments: the $(+,+)$ given to a
fermionic component (as for the matter hypermultiplets) or to a
scalar component, with the parities of all other fields fixed
automatically. Only the second choice leads to a non anomalous
theory and leaves the zero mode of the Higgs field massless. The
two Higgsinos, $(+,-)$ and $(-,+)$, are all paired to get a Dirac
mass at $(2n+1)/R$, $n=0,1,...$ (see Fig. \ref{fig:spectrum}).

\subsection{Residual symmetries after the orbifold projection}
\label{sec:symmetries}

The 5D supersymmetric gauge Lagrangian is fixed at this stage. The
symmetries that survive the orbifold projection other than gauge
and flavor symmetries are:
\begin{enumerate}
\item 5D supersymmetry with y-dependent transformation parameters
$\xi_1$, $\xi_2$ subject to the boundary conditions $(+,-)$ and
$(-,+)$ \cite{Barbieri:2002dm}. This implicitly assumes the
promotion of supersymmetry to a local symmetry, hence to
supergravity. We note that the scale of the supergravity couplings
need not be connected with the cut-off $\Lambda$ of Section
\ref{sec:LEP-paradox}.
\item A continuous R-symmetry with R-charges given in Table
\ref{tab:r-charges}, intact even after EWSB. The absence of any
A-terms or Majorana gaugino masses can be traced back to this
symmetry.
\begin{table}
\begin{center}
\begin{tabular}{|c||c|c|c|} \hline
$R$  & gauge $V$           & Higgs $H$     & matter $M$
\\ \hline
+2   &                     & $h^c$         &                          \\
+1   & $\tilde{\lambda} $  & $\tilde{h}^c$ & $\tilde{m}, \tilde{m}^c$ \\
0    & $A^\mu, A^c$        & $h$           & $m, m^c$                 \\
$-1$ & $\tilde{\lambda}^c$ & $\tilde{h}$   &
\\ \hline
\end{tabular}
\caption{Continuous $R$ charges for gauge, Higgs and matter
components. Here, $m$ represents $q, u, d, l, e$.}
\label{tab:r-charges}
\end{center}
\end{table}
\item A local $y$--parity under which any field transforms as
\begin{equation}
\varphi (y) \rightarrow \eta \, \varphi (\pi R /2 -y)
\label{eq:y-parity}
\end{equation}
where $\eta$ is the parity assignment at any one of the two
boundaries. Note that this cannot be extended to a global 5D
parity symmetry which includes the two boundaries since $Z_2
\times Z_2'$ is the most general discrete symmetry group on $S^1$
\cite{Barbieri:2002dm}. This symmetry is enough however to forbid
local mass terms for the hypermultiplets.
\end{enumerate}

These symmetries strongly constrain the form of the 5D (bulk)
Lagrangian, ${\cal L}_5$, but leave open the possibility of
suitable Lagrangian terms at the two boundaries, so that, for the
total Lagrangian
\begin{equation}
{\cal L} = {\cal L}_5 + \delta \left(y \right) {\cal L}_4 + \delta
\left( y-\frac{\pi R}{2}\right) {\cal L}_4'. \label{eq:lagrangian}
\end{equation}
Some of the terms in ${\cal L}_4$ and ${\cal L}_4'$ will in fact
be anyhow generated, subject only to the usual non-renormalization
properties of supersymmetry. One important fact to notice about
${\cal L}_4$ and ${\cal L}_4'$ is that they respect different
$N=1$ supersymmetries, associated to the parameters $\xi_1$ and
$\xi_2$, which vanish respectively at $y=\pi R/2$ and $y=0$. In
practice, to write down the most general ${\cal L}_4$ and ${\cal
L}_4'$ one employs the usual rules of 4D $N=1$ supersymmetry after
identification of the proper supermultiplets. In turn this
identification can be done by considering the 5D supersymmetry
transformation and setting there either $\xi_1$ or $\xi_2$ equal
to zero. If one looks at the Higgs hypermultiplet it is immediate
to see, anyhow, that the supermultiplets whose components have the
same orbifold parities and do not vanish at the boundaries are
$(h(+,+),\tilde h(+,-))$ and $(h^{\dagger}(+,+),\tilde h^c(-,+))$
respectively at $y=0$ and \mbox{$y=\pi R/2$}. This is what makes
possible to write down Yukawa couplings both for up and for down
quarks or for the leptons to a single Higgs field $h (+,+)$ and
still be consistent with (local) supersymmetry. The Yukawa
couplings for the up quarks are located at $y=0$ and the Yukawa
couplings for the down quarks and the leptons at $y=\pi R/2$
\cite{Barbieri:2001vh}.

Finally we note that ${\cal L}_4$ and ${\cal L}_4'$ can contain a
Fayet-Iliopoulos term associated with the hypercharge $U(1)$. We
shall come back to this possible term in Subsect.
\ref{sec:FI-term}.

\subsection{Gauge anomalies and hypermultiplet mass terms}

The boundary conditions, or the orbifolding, turn the vectorial 5D
Lagrangian into a chiral theory. This is obviously the case in the
pure gauge--matter sector since the orbifold projections select
chiral fermionic zero modes. It is also true however in the
gauge--Higgs sector in spite of parity conservation and of the
Dirac nature of all Higgsino masses. Some of the Kaluza Klein
vector bosons couple to vector currents and some others to axial
currents. Similarly some of the KK states of the gauge multiplet
$\varphi$ are scalars and some pseudoscalars. One wonders then if
gauge anomalies may appear localized on the boundaries
\cite{Arkani-Hamed:2001is,Scrucca:2002eb}.

The naive answer to this question turns out to be correct. To
ensure gauge invariance and the conservation of the corresponding
5D gauge current, it is enough that the fermionic zero modes,
after the orbifold projection, satisfy the usual 4D anomaly
cancellation condition \cite{Arkani-Hamed:2001is}.
Since the matter fermions are anomaly free and there are no
massless Higgsinos, the orbifold construction described above is
anomaly free. A qualification of this statement is necessary
however. Because of the Higgs sector, gauge invariance can be
maintained at the quantum level, but not, at the same time, the
local parity symmetry defined in Subsect. \ref{sec:symmetries}. In
particular there is no regularization that preserves both
symmetries \cite{Pilo:2002hu,Barbieri:2002ic}.

The breaking of the local $y$--parity makes it possible that there
be mass terms for the hypermultiplets. For the hypermultiplet of
components $(\psi,\psi^c,$ $\varphi,\varphi^c)$, the 5D mass term
consistent with the residual supersymmetry after the orbifold
projection is \cite{Barbieri:2002ic}
\begin{eqnarray}
{\cal L}_m &=&-\left(\psi m(y) \psi^c + \hc \right)
- M^2 \left(\abs{\varphi}^2+\abs{\varphi^c}^2 \right) \nonumber\\
 && -2M \left(\delta(y) + \delta
(y-\pi R/2)\right)\left(\abs{\varphi}^2-\abs{\varphi^c}^2\right)
\label{eq:lagrangian-mass}
\end{eqnarray}
irrespective of the specific boundary conditions for the different
components. Note the appearance of the boundary term. In the
formulation of the theory on a circle $S^1$, the mass term $m(y)$
has to satisfy $(-,-)$ boundary conditions to be also $Z_2 \times
Z_2'$ invariant. Furthermore, bulk supersymmetry implies that
$m(y)$ be piecewise constant in the four different patches of the
circle, hence
\begin{equation}
m(y)=M \eta(y) \, , \ \ \ \eta(y)=\left\{
\begin{array}{ll}
+1, & y \in (0,\pi R/2) \cup (\pi R, 3 \pi R/2) \\
-1, & y \in (\pi R/2, \pi R) \cup (3 \pi R/2, 2 \pi R)
\end{array} \right. \label{eq:eta-def}
\end{equation}
The effect of a mass term like (\ref{eq:lagrangian-mass}) on the
spectrum is discussed in Subsection \ref{sec:hyp-spectrum}. We
point out, however, that if these mass terms are vanishing at tree
level, the non-renormalization theorems guarantee that they can
only be renormalized by finite, negligibly small, non-local
corrections associated to the orbifold breaking of global
supersymmetry.

\subsection{The Fayet-Iliopoulos term} \label{sec:FI-term}

It was pointed out in Ref. \cite{Ghilencea:2001bw} that a
Fayet-Iliopoulos (FI) term on the boundaries is induced in the
model under examination by one loop corrections involving the
gauge coupling to the hypercharge $Y.$ At first this is not
surprising since a FI term in 4D is both gauge invariant and
globally supersymmetric. It is however also somewhat worrisome,
still in view of the 4D properties of a FI term. In 4D the FI term breaks
supersymmetry and/or the gauge symmetry in the vacuum, something
we would not like to happen in view of the previous
discussion. Furthermore, it is not gauge invariant in supergravity 
if the $U(1)$--charge of the gravitino vanishes \cite{Barbieri:1982ac,Ferrara:1983dh},
which is the case for $Y$. Finally the one loop FI term arises
only in presence of mixed $U(1)$--gravitational anomalies.

None of these unpleasant features necessarily survive in $5D$
\cite{Barbieri:2002ic}. In particular they are not shared by the
FI term in the model under consideration, which takes the form
\begin{equation}
\mathcal{L}_{\xi}=\xi\left(  \delta\left(  y\right)  \left(  X_{3}%
-\partial_{y}\Sigma\right)  +\delta\left(  y-\frac{\pi
R}{2}\right)  \left(
X_{3}+\partial_{y}\Sigma\right)  \right)  \label{fi1}%
\end{equation}
where $\Sigma$ is the real scalar in the vector hypermultiplet and
$X_{a}$ is the $SU(2)_{R}$ triplet of auxiliary fields of the 5D
vector multiplet \cite{Mirabelli:1997aj}. $X_{3}$ and $\Sigma$ intervene in the quadratic
Lagrangian without a mixed term
\begin{equation}
\mathcal{L}^{\left(  2\right)
}=\frac{1}{2}X_{3}^{2}+\frac{1}{2}\left(
\partial_{M}\Sigma \right) \left( \partial^{M}\Sigma \right)  \label{fi2}%
\end{equation}

For the purposes of this paper, it is important to observe that,
in the vacuum, from (\ref{fi1}),(\ref{fi2})
\begin{subequations}
\label{fi3}
\begin{align}
X_{3} &  =-\xi\left(  \delta\left(  y\right)  +\delta\left(  y-\frac{\pi R}%
{2}\right)  \right)  \\
\partial_{y}\Sigma &  =-\xi\left(  \delta\left(  y\right)  -\delta\left(
y-\frac{\pi R}{2}\right)  \right)
\end{align}
\end{subequations}
showing explicitly that the D--flatness conditions at both
boundaries $D=X_{3}-\partial_{y}\Sigma=0$,
$D^{\prime}=X_{3}+\partial_{y}\Sigma=0$ are satisfied. Note that,
on the $S^{1}$ circle, the vacuum form of $\Sigma$ is
$\left\langle \Sigma\right\rangle =-\frac{\xi}{2}\eta\left(
y\right)  $ with $\eta\left(  y\right)  $ as in
(\ref{eq:eta-def}). This amounts to a spontaneous breaking of the
local $y$--parity. In turn, after replacement of $\left( \ref{fi3}
\right)  $ in the interaction terms of the $X_{3}$ and $\Sigma$
fields with a generic hypermultiplet of hypercharge $Y$,
\begin{align}
\mathcal{L}_{\text{int}} &  =g_{Y}Y\left(
X_{3}-\partial_{y}\Sigma\right) \left(  \left|  \varphi\right|
^{2}-\left|  \varphi^{c}\right|  ^{2}\right)
\nonumber\\
&  -\left|  \left(  \partial_{y}-g_{Y}Y\Sigma\right)
\varphi\right| ^{2}-\left|  \left(
\partial_{y}+g_{Y}Y\Sigma\right)  \varphi^{c}\right|
^{2}\nonumber\\
&  +\psi^{c}\left(  \partial_{y}-g_{Y}Y\Sigma\right)  \psi+\text{h.c.}%
\end{align}
one obtains a supersymmetric mass term as in
(\ref{eq:lagrangian-mass}), with $M=-g_{Y}Y\xi /2$, where
$g_{Y}$ is 5D hypercharge coupling.

Once more we are led to consider a mass term for the
hypermultiplets. With a momentum cut-off $\Lambda$, the
radiatively generated FI term is
\begin{equation}
\xi=\frac{g_{Y}}{16\pi^{2}}\frac{\Lambda^{2}}{2}%
\end{equation}
which, in turn, translates itself into a mass term for a
hypermultiplet of
hypercharge $Y$%
\begin{equation}
M_{\xi}\left(  Y\right)  =-Y\frac{g_{Y}^{2}}{16\pi^{2}}\frac{\Lambda^{2}}%
{4}=-Y\frac{g^{\prime2}R}{32\pi}\Lambda^{2} \label{massafi}%
\end{equation}
where $g^{\prime}$ is the usual $U(1)$ coupling in 4D.

For $\Lambda R\simeq5$ this is a small mass, compared for example
with the one loop mass induced for the Higgs by the top loop
(\ref{eq:corr-m2-result}). One can nevertheless consider an
arbitrary value of $\xi,$ as done in Sect. \ref{sec:EWSB-FI}.
Finally it should be pointed out that this induced FI term has a
geometric interpretation in 5D supergravity, suggesting that its
renormalization vanishes beyond one loop and that, with a proper
regularization, the case $\xi=0$ is not unconceivable as coming
from a suitable more fundamental theory \cite{Barbieri:2002ic}.

\subsection{Hypermultiplet spectrum in presence of a mass term}
\label{sec:hyp-spectrum}

\begin{figure}
\centerline{\epsfxsize=.8\textwidth \epsfbox{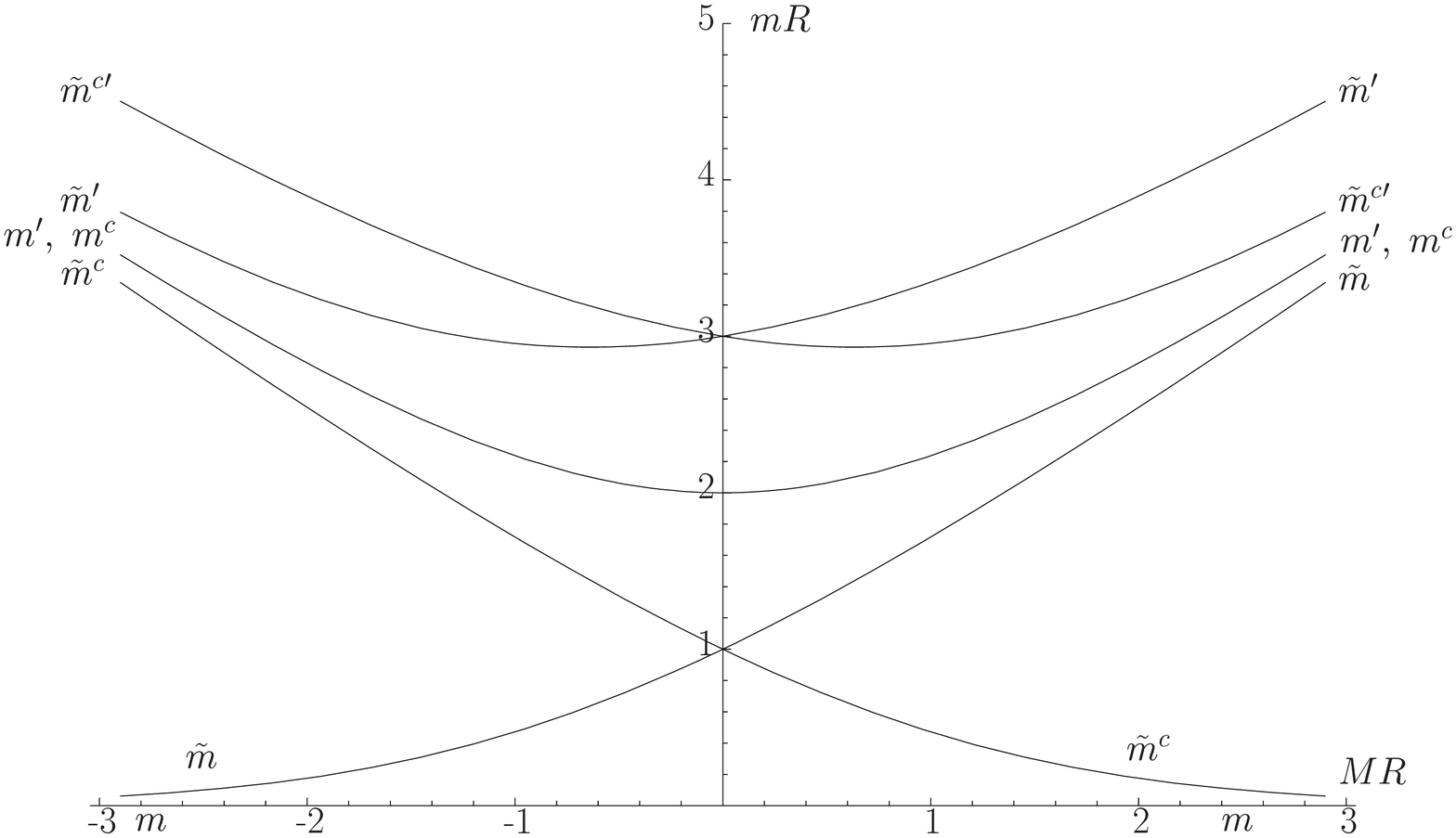}}
\caption{Spectrum of a matter hypermultiplet, in units of $1/R$,
as function of $MR$.} \label{fig:spettro}
\end{figure}

\begin{figure}
\centerline{\epsfxsize=.8\textwidth \epsfbox{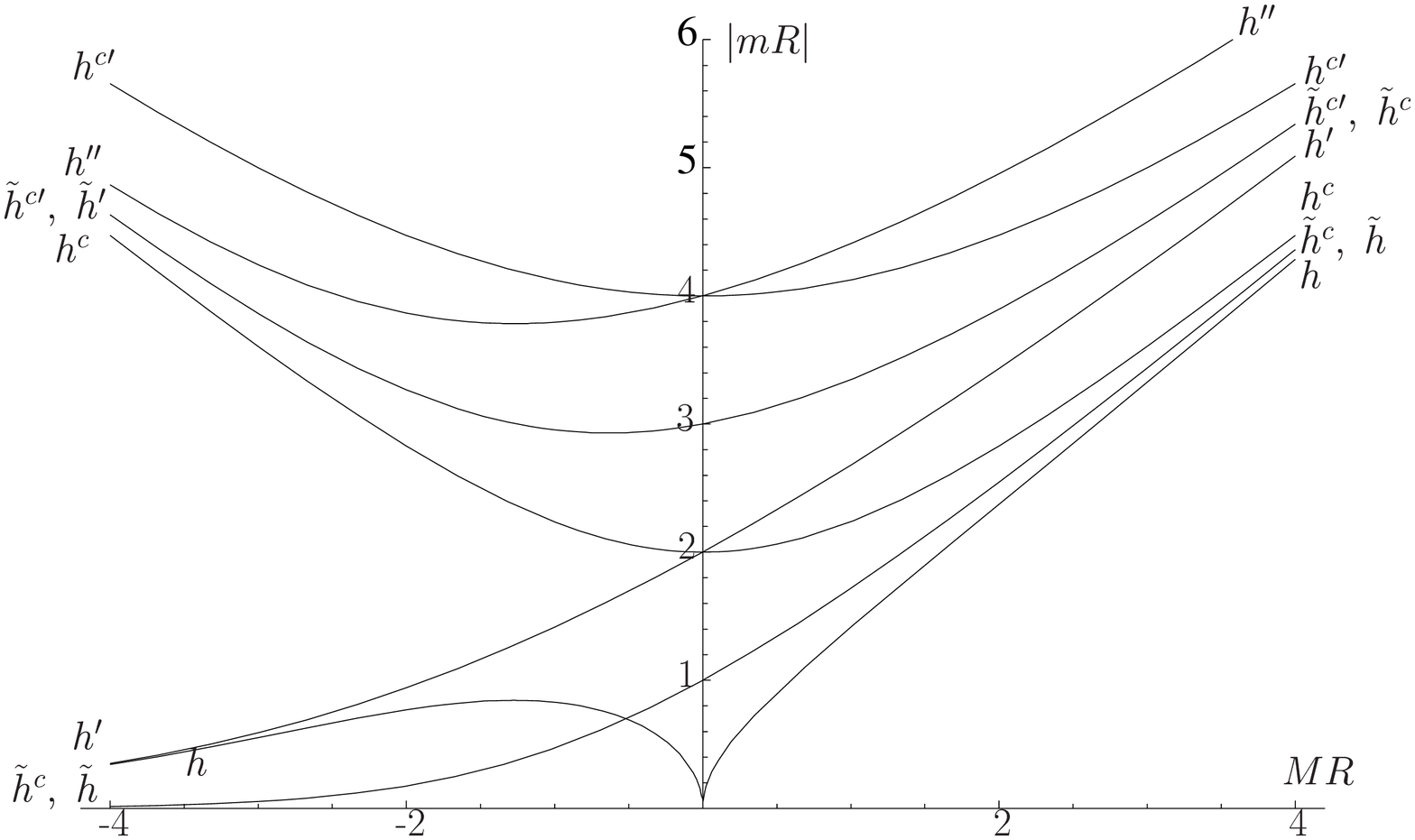}}
\caption{As in Fig. \ref{fig:spettro} for the Higgs
hypermultiplet.} \label{fig:spettro2}
\end{figure}

It is useful to summarize how the hypermultiplet spectrum of Fig.
\ref{fig:spectrum} is modified in presence of a mass term $M$ as
in $\left(\ref{eq:lagrangian-mass}\right)$. This spectrum is
worked out in Appendix \ref{sec:app-spectrum} both in the case of
matter-like and Higgs--like boundary conditions. The spectra in
the two cases are shown in Figs. \ref{fig:spettro},
\ref{fig:spettro2} respectively.
A few things are useful to note. In the large $MR$ limit a supersymmetric
spectrum is restored with bound states localized at the
boundaries. In Fig. \ref{fig:spettro2} the lightest state which
passes through zero at vanishing $MR$ is the Higgs, with the cusp
at $MR=0$ reflecting the change of sign of the squared mass, 
being $\left(  m_{h}R\right)  ^{2}\simeq\frac{4}{\pi}MR$
at $\left| MR \right| \ll 1$.

\section{Electroweak symmetry breaking in detail}\label{sec:EWSB}

The purpose of this Section is to study in detail the possible
effects of hypermultiplet masses on EWSB. In this paper we consider
the case that these masses do not exceed $1/R$, leaving the
exploration of the alternative possibility to a future publication. As
seen in Sect. \ref{sec:model}, moderate values of $MR$ are consistent
with radiative corrections.  Other than the modification of the
spectrum, hypermultiplet masses have 3 types of effects on the EWSB:

\begin{enumerate}
\item  a mass term for the top $Q$ or $U$ hypermultiplets changes the relation
between the top mass and the top-Higgs coupling, crucial in EWSB;

\item  a mass term for the top $Q$ or $U$ hypermultiplets influences the
one loop Higgs mass or the complete one loop effective potential;

\item  a mass term for the Higgs hypermultiplet gives a tree level mass to the
zero mode Higgs field (see Fig. \ref{fig:spettro2}), which feeds
directly in the effective potential.
\end{enumerate}

\subsection{The top mass and the top Yukawa coupling}

The relation between the top mass and the top--Higgs coupling
$y_{t}$ is obtained by solving the equation of motion for the
lowest mode of the fermions in the top hypermultiplets $Q$ and
$U$, coupled by (\ref{eq:yukawa-term}) at the $y=0$ boundary. The
interaction (\ref{eq:yukawa-term}) is itself a localized mass term
when the Higgs scalar is replaced by the vacuum expectation value
$v$. This is done in Appendix \ref{sec:app-spectrum}. The result
can be expressed as
\begin{equation}
y_t=\widehat{y}_t \eta^U_0 \eta^Q_0
 \eta^h_0
\label{eq:uaiti2}
\end{equation}
where
\begin{align}
\widehat{y}_t &= \frac{\lambda_t}{\left( 2 \pi R \right)^{3/2}}=\frac{2
m_t}{\pi v}\frac{1}{\sqrt{\omega_-^U \omega_-^Q}}\\
\omega_{\pm}^i &= k_i R \coth \left( \frac{k_i \pi R}{2}\right) \pm
M_i R, \ \ \ \ \ \ \ i=U,Q\\
k_i&=\left(M_i^2 -m_t^2 \right)^{1/2}
\end{align}
and $\eta^{U,Q,h}_0$ are the wave functions for the lightest $U,Q,h$
modes at $y=0$, normalized to $\int_{0}^{2 \pi R} \textrm{d}y \left|
\eta^i (y) \right|^2=2 \pi R$ and given in Appendix
\ref{sec:app-spectrum}. At $M_U=M_Q=M_h=0$, (\ref{eq:uaiti2}) reduces
to
\begin{equation}
y_t=\frac{m_t}{v} \frac{2 \sin \left(\pi R m_t\right)}{\pi R m_t + \sin \left(\pi R m_t\right)}
\end{equation}
Note that in the limit $ M_h=0$, $ \left| M_{U,Q} \right| \gg 1/R \gg m_t$, $y_t$
reduces to the standard value, $m_t/v$, no matter what the sign of $M$
is. For positive $M$, when the top wave function and the Yukawa
coupling are localized at opposite boundaries, this is due to a
compensating increase of $\widehat{y}_t$, which is directly related to
the fundamental coupling in the Lagrangian. When $y_t$ and
$\widehat{y}_t$ differ significantly, it is $\widehat{y}_t$ that
enters into (\ref{eq:rel5D-yt-r-lambda})-(\ref{eq:rel4D-yt-r-lambda})
to determine the point of saturation of perturbation theory.

\subsection{One loop Higgs effective potential for arbitrary $M_{U}$, $M_{Q}$}

The one loop Higgs mass $\left(\ref{eq:corr-m2-result}\right)  $
from the diagrams of Fig. \ref{fig:1loop-diagrams} gets corrected
by the presence of $M_{U}$ and $M_{Q}$. This is in fact also true
for the entire one loop effective potential which has to be
computed anyhow because of the large correction to the quartic
coupling and because of the higher order terms in $\left(
vR\right)  ^{2}$ which may be important insofar as $R$ is not
determined.

The calculation described in Sect. \ref{sec:LEP-paradox}
immediately generalizes to the massive case in terms of the
propagators in presence of masses. Considering,
as in $\left(\ref{eq:corr-m2}\right)$, the mixed propagators at
$y=y^{\prime}=0$ the effective potential due to top--stop exchanges is
\begin{align}
V_{t}\left(  h;M_{U},M_{Q} \right) = & \frac{N_c}{2}
\sum_{N=1}^{\infty} 
\int \frac{d^4 p}{\left( 2 \pi \right)^4} \frac{ (-1)^{N+1}}{N}
\left(\frac{\lambda_t \, h \, \eta_0^h}{2 \sqrt{2 \pi R}}\right)^{2N} \nonumber \\
& \left\{  \left[ G_{\varphi}^{U} \left( p,0 \right) G_{F}^{Q} \left( p,0
   \right)\right]^N 
 +  \left[ G_{\varphi}^{Q} \left( p, 0 \right) G_{F}^{U} \left( p, 0 \right)\right]^N 
 -   \, 2 \left[ G_{\psi}^{U} \left( p,0\right)
  G_{\psi}^{Q} \left( p,0 \right) \right]^N \right\} \label{potenzialeeffettivo}
\end{align}
where $G^{U,Q}_i\left(p,y\right)=G_i\left(p,y;M_{U,Q}\right)$ with $i=\phi, \, F,
\, \psi$. The propagators $G_i\left(p,y;M\right)$ are given in
Appendix \ref{sec:app-propag}, while the wave function of the Higgs zero mode
 $\eta_0^h$ is given in Appendix \ref{sec:app-spectrum}.

\subsection{Electroweak symmetry breaking in presence of a FI
term}\label{sec:EWSB-FI}

As shown in Sect. \ref{sec:FI-term}, a FI term is equivalent for
any hypermultiplet of hypercharge $Y$ to a mass term, which we
parameterize in terms of a dimensionless variable $a$ as%
\begin{equation}
M\left(  Y\right)  =\frac{a}{R}Y, \label{eq:def-a}
\end{equation}
to be inserted in (\ref{eq:lagrangian-mass},\ref{eq:eta-def}).

In presence of these masses the potential we consider to determine
the VEV of the Higgs field is%
\begin{align}
V\left(  h;R,a\right)   &  =m^{2} \left( M\left(  1/2\right) \right)
h^{2}+\frac{21\xi\left(
3\right)  }{16\pi^{4}}\frac{g^{2}}{R^{2}}h^{2}\nonumber\\
&  +\frac{g^{2}+g^{\prime2}}{8}h^{4}+V_{t}\left(  h;M\left(
-2/3\right) ,M\left(  1/6\right)  \right)
\label{eq:pot-y-dependent}
\end{align}

Other than the standard tree-level quartic coupling and the one
loop contribution from the top-stop exchanges, eq. $\left(
\ref{potenzialeeffettivo}\right)$, the potential includes the tree level mass
$m(M)$ computed in Sect. \ref{sec:hyp-spectrum} and Appendix
\ref{sec:app-spectrum} (first term on the r.h.s. of
(\ref{eq:pot-y-dependent})) and a one loop mass term from the KK tower
of the $SU(2)$ gauge multiplets (second term on the r.h.s. of 
(\ref{eq:pot-y-dependent}))\cite{Antoniadis:1998sd}.
\begin{figure}[!t]
\centerline{\epsfxsize=.49\textwidth \epsfbox{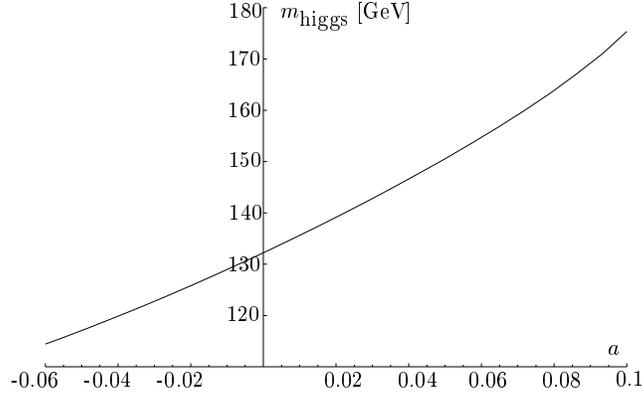}}
\caption{Higgs mass as function of the dimensionless parameter
$a$, eq. (\ref{eq:def-a}).} \label{fig:mhiggs-a}
\end{figure}
\begin{figure}[!h]
\centerline{\epsfxsize=.50\textwidth \epsfbox{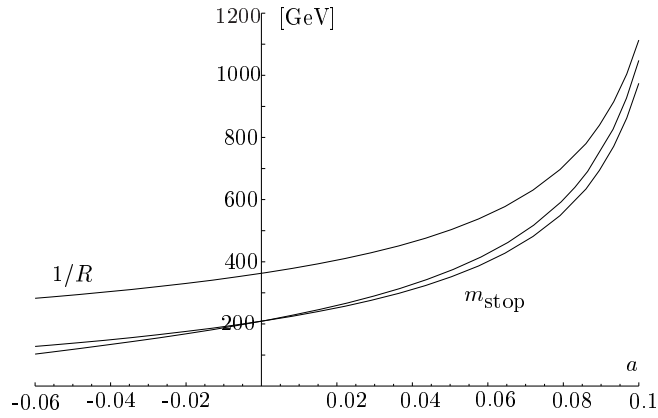}}
\caption{Compactification scale and lightest stop masses as
functions of $a$, eq. (\ref{eq:def-a}).} \label{fig:mstop-a}
\end{figure}
\begin{figure}[!h]
\centerline{\epsfxsize=.49\textwidth \epsfbox{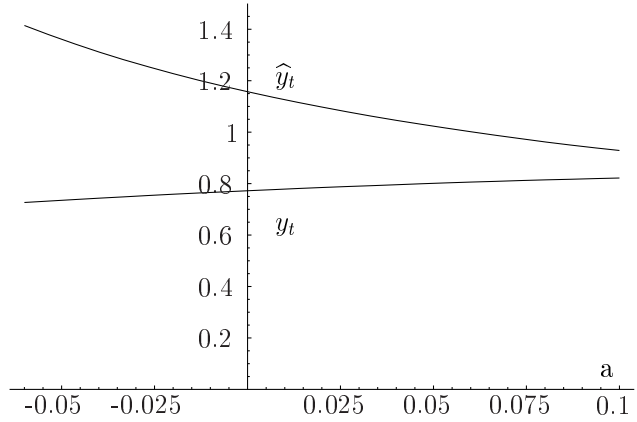}}
\caption{Top--Higgs coupling ($y_t$) and $\widehat y_t = \lambda_t
/(2 \pi R)^{3/2}$ as functions of $a$, eq. (\ref{eq:def-a}).} \label{fig:uaiti-a}
\end{figure}

Imposing the occurrence of the minimum at $h=v=175$ GeV determines
the Higgs mass $m_{h}$ and $1/R$, together with the entire
spectrum, as functions of $a$. The Higgs mass is shown in Fig.
\ref{fig:mhiggs-a}. The lightest stops, which are non degenerate when
$a \neq 0$ because $M_U \neq M_Q$, occurs in two chiralities. Their
mass difference depends on the parameter $a$ and is about $70 \, \gev$
at $a=0.1$. The stop masses together
with $1/R$ are shown in Fig. \ref{fig:mstop-a}. These figures refine those of Ref.
\cite{Barbieri:2001cz}.

The sharp increase of $1/R$ with $a$ is due to an increasingly
precise accidental cancellation (at about $10\%$ level for
$a=0.1$) between the positive tree level squared mass in
(\ref{eq:pot-y-dependent}) and the negative contribution from the
top-stop loop. Note that the estimate of the radiatively induced
FI term in $\left( \ref{massafi}\right)  $ corresponds to a
negligibly small $a=5\cdot10^{-3}$ \cite{Barbieri:2001cz}.

Through $R$, $M_{U}$ and $M_{Q}$, also the Higgs--top coupling
acquires a dependence on $a$, determined in eq. $\left(
\ref{eq:uaiti2}\right)  $ and shown in Fig. \ref{fig:uaiti-a}.
Note that the top Yukawa coupling $y_t$ is reduced from the
Standard Model value by about $10\%$ due to the localization of
the interaction at the boundary.

\subsection{Electroweak symmetry breaking with sizeable $M_{U}=M_{Q}$}

As we have seen, the mass terms from the FI term have to be small.
Their effect can however be significant due to a possible
cancellation occurring in the Higgs potential between the tree
level Higgs squared mass and the radiatively induced effect. Here
we consider the possible effects of direct masses for the $U,Q$
hypermultiplets, taking $M_{U}=M_{Q}=M$ for simplicity. At the
same time, again for simplicity, we set the FI term, or the $a$
parameter, to zero.

\begin{figure}[!thb]
\centerline{\epsfxsize=.5\textwidth \epsfbox{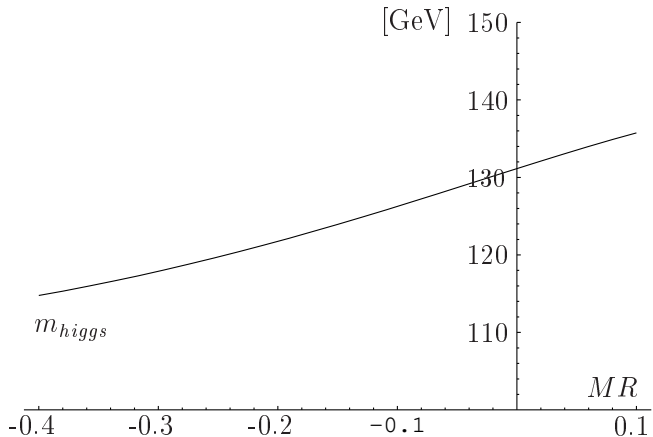}}
\caption{Higgs mass as function of $MR$.} \label{fig:mhiggs-MuMq}
\end{figure}

\begin{figure}[!ht]
\centerline{\epsfxsize=.5\textwidth \epsfbox{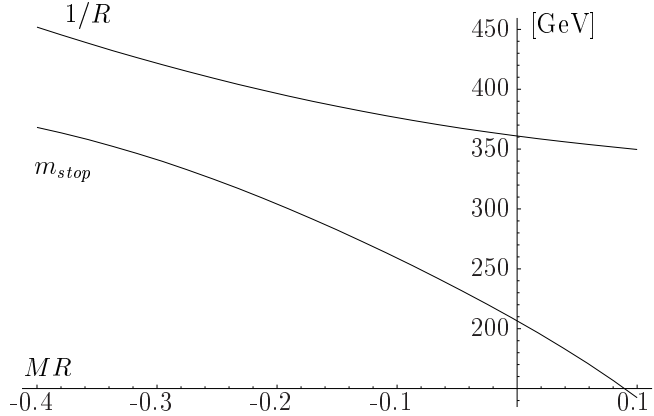}}
\caption{Stop mass and 1/R as functions of $MR$.} \label{fig:mstop-MuMq}
\end{figure}

\begin{figure}[!ht]
\centerline{\epsfxsize=.5\textwidth \epsfbox{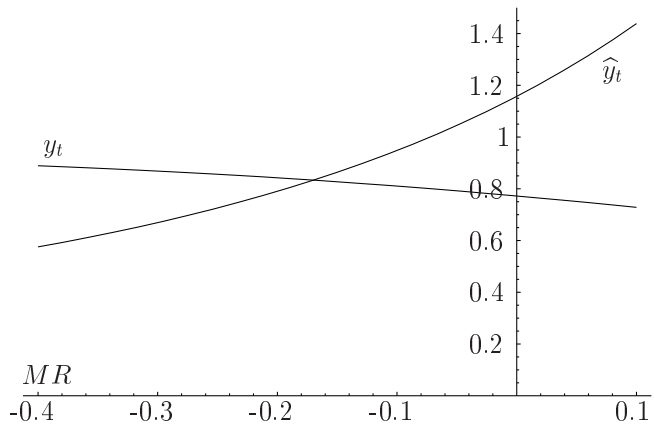}}
\caption{Top-Higgs coupling $y_t$ and $\widehat y_t = \lambda_t/(2 \pi
  R)^{3/2}$ as functions of $MR$.} \label{fig:uaiti-MuMq}
\end{figure}

Proceeding as in the previous section, the Higgs potential we consider is%
\begin{equation}
V\left(  h;R,M\right)  =\frac{21\xi\left(  3\right)  }{16\pi^{4}}\frac{g^{2}%
}{R^{2}}h^{2}+\frac{g^{2}+g^{\prime2}}{8}h^{4}+V_{t}\left(
h;M,M\right)
\end{equation}
whose minimization determines $m_{h}$ and $R$ as functions of $M.$
The Higgs mass is shown in Fig. \ref{fig:mhiggs-MuMq} for $-0.4 \leq
MR  \leq 0.1$. The reason
for interrupting $MR$ below $0.1$ is that the mass of lightest stops\footnote{The lightest stops now come in two
 degenerate chiralities.} falls below the experimental lower bound of about \mbox{$150$
{\gev}}, as shown in Fig. \ref{fig:mstop-MuMq}. 
For $MR$ below $-0.4$, instead, it is the Higgs boson which becomes too light. This
result, however, could not persist for $MR < -1$, where higher-loop gauge corrections become
important \cite{Marti:2002ar}. This case will be analyzed elsewhere.
In the interval $-0.4 \leq  MR  \leq 0.1$, both $1/R$ and $m_{\tilde
  t}$ have a non negligible dependence an $M$, as shown in
Fig. \ref{fig:mstop-MuMq}. The degeneracy between the two lightest stop masses would
be resolved by taking $M_U \neq M_Q$. The Top-Higgs couplings $y_t$
and $\widehat y_t$ in this case are shown in Fig. \ref{fig:uaiti-MuMq}.

\section{Spectrum and phenomenological
implications}\label{sec:phenomenology}

In absence of hypermultiplet mass terms, the value of the
compactification scale and the spectrum of the lightest particles
is given in Table \ref{tab:modes-masses} with an error that
estimates the uncertainties due to the presence of the extra
couplings and operators mentioned in Sect. \ref{sec:rel-cutoff} \cite{Barbieri:2001vh}.
\begin{table}
\begin{center}
\begin{tabular}{|c||c|c|} 
\hline
 & A & B \\
\hline
\hline
$1/R$                        & $360 \pm 70 $  & $300 \div 1000 $ \\
\hline
$h$                                  & $133 \pm 10 $  & Fig. \ref{fig:mhiggs-1suR} \\
\hline
$\tilde t_1 , \, \tilde u_1$          & $210 \pm 20 $  & $1/R (1 \pm 8\%) - m_t$ \\
\hline
$\chi^{\pm}, \, \chi^0,$              &                   & \\
$\tilde g, \, \tilde q, \, \tilde l$ & $360 \pm 70 $  & $1/R (1 \pm 20\%)$ \\
\hline
$\tilde t_2, \, \tilde u_2$           & $540 \pm 30 $  & $1/R (1 \pm 8\%) + m_t$ \\
\hline
$A_1,q_1,l_1,h_1$                    & $720 \pm 140 $  & $2/R (1 \pm 20\%)$ \\
\hline
\end{tabular}
\caption{The particle spectrum and $1/R$ in absence of any mass term
  (Column A) and in presence of a FI term (Column B). All entries are in GeV.} \label{tab:modes-masses}
\end{center}
\end{table}
By letting the mass terms vary in a moderate range, well consistent 
with radiative corrections, the main deviation from the massless 
case is due to a possible mass term for the Higgs hypermultiplet which can partially
counteract the top-stop radiative corrections that trigger
EWSB. This can in turn drive up the compactification scale and,
consequently, the entire spectrum.

\begin{figure}[bht]
\centerline{\epsfxsize=.5\textwidth \epsfbox{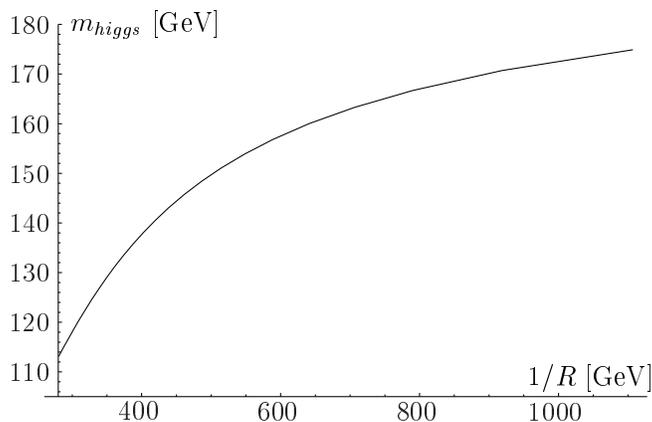}}
\caption{Higgs mass versus $1/R$ in presence of a Fayet-Iliopoulos
term.} \label{fig:mhiggs-1suR}
\end{figure}

In Sect. \ref{sec:EWSB-FI} we have explicitly discussed the
effects of a FI term, which is a particular example of this case.
The entire spectrum becomes therefore effectively determined by
$1/R$ in the range of Fig. \ref{fig:mstop-a}, $300 \gev \lesssim
R^{-1} \lesssim 1000 \gev$. The dependence of $m_{h}$ on $1/R$ is
shown in Fig. \ref{fig:mhiggs-1suR} obtained from Fig.
\ref{fig:mhiggs-a}--\ref{fig:mstop-a}, whereas the masses of the
other particles is again given in Table \ref{tab:modes-masses}.
Note that the lightest stop $\widetilde{t}_{1}$ is the Lightest
Supersymmetric Particle (LSP), except possibly for large values of
$1/R$ where the corrections due to kinetic terms localized on the
boundaries, giving rise to the main uncertainty indicated in Table
\ref{tab:modes-masses}, could reverse the order with any of the
other superpartners at $1/R.$ Unless an explicit violation of the
$U(1)_{R}$--symmetry were introduced at the boundaries, the LSP
would be stable.
A moderate effect could also arise from an explicit mass term for the
top hypermultiplets, as shown in Fig. \ref{fig:mstop-MuMq}.

\subsection{Phenomenological implications}

Except for the large, somewhat fine tuned values of $1/R,$ the
Higgs boson is below the $WW$ threshold, with a preferred mass in
the 130 GeV range. It has SM-like couplings to $b\overline{b}$ and
$\tau\overline{\tau}$ and $WWh,$ $ZZh$ gauge couplings. It could
therefore be looked at in associated production of $Wh$ or $Zh$,
followed by $b\overline{b}$ and $\tau\overline {\tau}$ decays. We
have already mentioned the deviation of the top Yukawa coupling
from the SM value (see Figs. \ref{fig:uaiti-a},
\ref{fig:uaiti-MuMq}). More important for the possible discovery
in a hadron collider is the suppression of the Higgs--gluon--gluon
squared coupling, ranging from $10 \%$ to $60 \%$ relative to the
SM value as $1/R$ increases from 300 to about 700 GeV, where the
$WW$ threshold is crossed \cite{Cacciapaglia:2001nz}.

A main feature of the model is that the two degenerate light stops
are the LSP and are stable if $U(1)_{R}$ is exact. Their mass is
approximately $\left( 1/R-m_{t}\right)  $ with a lowest preferred
value in the 200 GeV range. In such a low range value the
super-hadrons $T^{+}=\widetilde{t}_{1}\overline{d}$ and
$T^{0}=\widetilde {t}_{1}\overline{u}$ and their charge conjugates
$T^{-},\overline{T}^{0}$ could easily be detected at the Tevatron
Run II as stable particles, since their possible decay into one
another is slow enough to let them both cross the detector. $T^{\pm}$
could appear as a stiff\ charge track with little hadron
calorimeter activity, hitting the muon chambers and
distinguishable from a muon via d$E/$d$x$ and time-of-flight. The
neutral states, on the contrary, could be identified as missing
energy since they could traverse the detector with little
interaction. The cross section for the pair production at the
Tevatron of the stops with a 200 GeV mass is $0.6$ pb
\cite{Chertok:2001ja}.

The heavier supersymmetric particles in Table 2 could be looked at
through their chain decay into the LSP. Similarly the discovery of
the first states at $2/R$ of the KK tower of SM particles (heavy
quarks, leptons with their mirror partners, heavy gauge and Higgs
bosons) would be strong evidence for the picture of EWSB described
in this paper. Note that (discrete) momentum conservation
in the 5th dimension forbids unsuppressed gauge couplings of the
heavy gauge bosons to the standard fermions.

\appendix

\section*{Appendices}

\section{Spectrum} \label{sec:app-spectrum}

In this appendix we calculate the KK spectrum of Higgs and matter
hypermultiplets in presence of a mass term $M$ as in $\left(
\ref{eq:lagrangian-mass}\right)  $.

Let $\left(  \varphi,\psi,F\right)$ ,$\left(  \varphi%
^{c},\psi^{c},F^{c}\right)  $ be either a Higgs or a matter
hypermultiplet. In presence of a mass term as in $\left(
\ref{eq:lagrangian-mass}\right)  $ the Lagrangian upon eliminating the
F-terms is:%
\begin{align}
\mathcal{L} &  =\abs{\partial_{M}\varphi}^2  +\abs{\partial_{M}\varphi^{c}}^2
  +i\psi\sigma^{\mu}\partial_{\mu}\overline{\psi}+i\psi^{c}%
\sigma^{\mu}\partial_{\mu}\overline{\psi}^{c}+\psi^{c}\partial_{y}%
\psi+\overline{\psi}^{c}\partial_{y}\overline{\psi}\nonumber\\
&  -M^{2}\left(  \left|  \varphi\right|  ^{2}+\left|  \varphi%
^{c}\right|  ^{2}\right)  -2M\left(  \delta\left(  y\right)
+\delta\left(  y-\frac{\pi R}{2}\right)  \right)  \left(  \left|  \varphi
\right|  ^{2}-\left|  \varphi^{c}\right|  ^{2}\right)  \nonumber\\
&  -M\eta\left(  y\right)  \left(  \psi\psi^{c}+\overline{\psi}%
\;\overline{\psi}^{c}\right)  \label{eq:lagrangiana}%
\end{align}
where $M=0,1,2,3,5$ while $\mu=0,1,2,3.$

Thus the equations of motion are%
\begin{subequations}
\label{eq:equazionidelmoto}
\begin{align}
&  \left[  \partial_{M}\partial^{M}+M^{2}+2M\left(  \delta\left(
y\right)  +\delta\left(  y-\frac{\pi R}{2}\right)  \right)  \right]
\varphi=0\\
&  \left[  \partial_{M}\partial^{M}+M^{2}-2M\left(  \delta\left(
y\right)  +\delta\left(  y-\frac{\pi R}{2}\right)  \right)  \right]
\varphi^{c}=0\\
&  \left[  \partial_{M}\partial^{M}+M^{2}+2M\left(  \delta\left(
y\right)  -\delta\left(  y-\frac{\pi R}{2}\right)  \right)  \right]  \psi
=0\\
&  \left[  \partial_{M}\partial^{M}+M^{2}-2M\left(  \delta\left(
y\right)  -\delta\left(  y-\frac{\pi R}{2}\right)  \right)  \right]  \psi
^{c}=0%
\end{align}
\end{subequations}

Equations $\left(  \ref{eq:equazionidelmoto}\right)  $ must be
solved imposing the proper boundary conditions to the fields
$\varphi$, $\varphi^{c}$, $\psi$, $\psi^{c}$. Note that the delta
functions in the left-hand side of equations $\left(
\ref{eq:equazionidelmoto}\right)  $ are both present only if the
field under consideration has $\left(  +,+\right)  $ parity under
$Z_{2}\times Z_{2}^{\prime}$ symmetry. In all other cases the wave
function vanishes at $y=0$ and/or $y=\frac{\pi R}{2}$ and the
delta functions in the corresponding points are irrelevant.

\subsection{Matter hypermultiplets}

If we consider a matter hypermultiplet, then $\left(  \ref{eq:equazionidelmoto}%
\right)  $ become%
\begin{subequations}
\label{eq:equazionimotomateria}
\begin{align}
&  \left[  \partial_{M}\partial^{M}+M^{2}+2M\left(  \delta\left(
y\right)  -\delta\left(  y-\frac{\pi R}{2}\right)  \right)  \right]
\psi=0 \label{eq:equazionemodozeromateria}\\
&  \left[  \partial_{M}\partial^{M}+M^{2}\right]  \psi^{c}=0\\
&  \left[  \partial_{M}\partial^{M}+M^{2}+2M\delta\left(  y\right)
\right]  \varphi=0\\
&  \left[  \partial_{M}\partial^{M}+M^{2}-2M\delta\left(  y-\frac{\pi
R}{2}\right)  \right]  \varphi^{c}=0%
\end{align}
\end{subequations}

Taking for the wave functions the following form%
\begin{subequations}
\begin{align}
&\psi\left(  x,y\right) = \left\{
\begin{array}
[l]{l}%
\widetilde{\psi}\left(  x\right)  \left[  A_{\psi}\sin k\left(  y-\frac{\pi
R}{2}\right)  +B_{\psi}\cos k\left(  y-\frac{\pi R}{2}\right)  \right]  \\
\widetilde{\psi}\left(  x\right)  \left[  -A_{\psi}\sin k\left(  y-\frac{\pi
R}{2}\right)  +B_{\psi}\cos k\left(  y-\frac{\pi R}{2}\right)  \right]
\end{array}
\right.
\begin{array}
[c]{l}%
y\in\left(  0,\frac{\pi R}{2}\right)  \\
y\in\left(  \frac{\pi R}{2},\pi R\right)
\end{array}
\\
&\psi^{c}\left(  x,y\right)   = \widetilde{\psi}^{c}\left(  x\right)
A_{\psi^{c}}\sin ky \hspace{5.97cm} y \in\left(  0,\pi R\right)  \\
&\varphi\left(  x,y\right)    =\widetilde{\varphi}\left(  x\right)
A_{\varphi}\sin k\left(  y-\frac{\pi R}{2}\right) \hspace{4.62cm}  y  \in\left(  0,\pi
R\right)  \\
&\varphi^{c}\left(  x,y\right)    =\widetilde{\varphi}^{c}\left(  x\right)
\left\{
\begin{array}
[l]{l}%
A_{\varphi^{c}}\sin ky\\
B_{\varphi^{c}}\sin k\left(  \pi R-y\right)
\end{array}
\right.
\hspace{3.6cm}
\begin{array}
[l]{l}%
y\in\left(  0,\frac{\pi R}{2}\right)  \\
y\in\left(  \frac{\pi R}{2},\pi R\right)
\end{array}
\end{align}
\label{eq:funzionid'onda}
\end{subequations}
the mass of every field is given by $m^{2}=M^{2}+k^{2}$ where $k$ is
constrained by equations $\left(  \ref{eq:equazionimotomateria}\right)  .$
Imposing the proper conditions on the wave functions and their first
derivatives on the boundary we get the following equations for $k$%
\begin{subequations}
\label{eq:equazionimateria}
\begin{align}
\psi\left(  +,+\right)   &  \Longrightarrow\left(  k^{2}+M^{2}\right)
\sin\frac{k\pi R}{2}=0\\
\psi^{c}\left(  -,-\right)   &  \Longrightarrow\sin\frac{k\pi R}%
{2}=0\\
\varphi\left(  +,-\right)   &  \Longrightarrow\tan\frac{k\pi R}{2}=-\frac
{k}{M}\\
\varphi^{c}\left(  -,+\right)   &  \Longrightarrow\tan\frac{k\pi R}{2}%
=\frac{k}{M}%
\end{align}
\end{subequations}
A few things are worthy noticing:

\begin{enumerate}
\item  One can get the equations for the bound states by analytical continuation, setting $k%
=i\rho$ in $\left(  \ref{eq:equazionimateria}\right)  $

\item  The bound state $\psi\left(  +,+\right)  $ is massless for every value
of $M$, while the excited states have masses $\left(
m_{\psi}^{2}\right) _{n}=M^{2}+\left(\frac{2n}{R}\right)^2, \
n=1,2,\ldots$

\item  The equation for $\psi^{c}\left(  -,-\right)  $ is unaffected by the
presence of $M$ because of the vanishing of the wave function at
$y=0,\frac{\pi R}{2}$
\end{enumerate}

\subsection{Higgs hypermultiplet}

If we consider a Higgs hypermultiplet, then $\left(  \ref{eq:equazionidelmoto}%
\right)  $ become%
\begin{subequations}
\label{eq:equazionimotohiggs}
\begin{align}
&  \left[  \partial_{M}\partial^{M}+M^{2}+2M\left(  \delta\left(
y\right)  +\delta\left(  y-\frac{\pi R}{2}\right)  \right)  \right]
h=0 \label{eq:equazionemotohiggsscalare}\\
&  \left[  \partial_{M}\partial^{M}+M^{2}\right]  h^{c}=0\\
&  \left[  \partial_{M}\partial^{M}+M^{2}+2M\delta\left(  y\right)
\right]  \lambda=0\\
&  \left[  \partial_{M}\partial^{M}+M^{2}+2M\delta\left(  y-\frac{\pi
R}{2}\right)  \right]  \lambda^{c}=0%
\end{align}
\end{subequations}
With the same procedure of the matter case one gets the equations%
\begin{subequations}
\label{eq:equazionihiggs}
\begin{align}
h\left(  +,+\right)   &  \Longrightarrow\tan\frac{k\pi R}{2}=\frac{2kM}%
{k^{2}-M^{2}} \label{eq:equazionehiggsscalare}\\
h^{c}\left(  -,-\right)   &  \Longrightarrow\sin\frac{k\pi R}{2}=0\\
\lambda\left(  +,-\right)   &  \Longrightarrow\tan\frac{k\pi R}{2}=-\frac
{k}{M}\\
\lambda^{c}\left(  -,+\right)   &  \Longrightarrow\tan\frac{k\pi R}{2}%
=-\frac{k}{M}
\end{align}
\end{subequations}
Note that equation (\ref{eq:equazionehiggsscalare}) for the bound state of the $h$ field leads to a
negative squared mass if $M<0.$

Solving $\left(  \ref{eq:equazionemotohiggsscalare}\right)  $ for the zero mode of
the Higgs scalar we find for the wave function normalized to $\int_{0}^{2 \pi R} \textrm{d}y \left|
 h^{\left(0\right)} \left(  y\right) \right|^2=1$%
\begin{equation}
h^{\left(  0\right)  }\left(  y\right)  =\frac{-k\cos k\left(  y-\frac{\pi
R}{2}\right)  +M\sin k\left(  y-\frac{\pi R}{2}\right)  }{\sqrt{2M+\left(
k^{2}+M^{2}\right)  \pi R-2M\cos k\pi R+\left(  k-\frac{M^{2}}{k}\right)  \sin
k\pi R}}\label{eq:funzioned'ondahiggs}%
\end{equation}
with $k$ the solution of $\left( \ref{eq:equazionehiggsscalare}
\right)  .$ Expression (\ref{eq:funzioned'ondahiggs}) is valid for $y \in \left[0,\pi R/2 \right)$. Note that if $M\rightarrow 0,$ then $h^{\left(
0\right)  }\left(  y\right)  \rightarrow\left(  2\pi R\right)
^{-1/2}.$

\subsection{Spectrum in presence of a VEV for the Higgs field}

If we have a top quark hypermultiplet, then the top Yukawa coupling%
\begin{equation}
\mathcal{L}_{Y}=\frac{\lambda_{t}}{2}\left[  \delta\left(  y\right)
+\delta\left(  y-\pi R\right)  \right]  \int\text{d}^{2}\theta\widehat
{h}\,\widehat{Q}\,\widehat{U} + \textrm{h.c.}%
\end{equation}
leads to a mass term when we replace the Higgs zero mode
$h^{\left(  0\right) }$ with its VEV $v.$ To calculate the
spectrum in presence of such a term it is convenient to rewrite
the Lagrangian $\left(  \ref{eq:lagrangiana}\right) $ without
eliminating the $F$ auxiliary fields and use the following vectors
\begin{equation}
\begin{array}{ccc}
X=\left(\begin{array}{c} \varphi \\ F^{c \dagger} \end{array} \right)
& Y=\left(\begin{array}{c} \varphi^{c} \\ F^{\dagger}
\end{array} \right) & Z=\left(\begin{array}{c} \psi \\ \bar{\psi^c}
 \end{array} \right)
\end{array}
 \label{eq:def-vect-lagr}
\end{equation}
Then $\left( \ref{eq:lagrangiana}\right)  $ becomes%
\begin{align}
\mathcal{L} &  =\left(
\begin{array}
[c]{cc}%
X_{U,Q}^{\dagger} & Y_{Q,U}^{\dagger} %
\end{array}
\right)  M_{B}\left(
\begin{array}
[c]{c}%
X_{U,Q}\\
Y_{Q,U}
\end{array}
\right) \nonumber \\
&  +\left(
\begin{array}
[c]{cc}%
\overline{Z}_{U,Q} & Z^t_{Q,U}
\end{array}
\right)  M_{F}\left(
\begin{array}
[c]{c}%
Z_{U,Q}\\
\overline{Z}^t_{Q,U}\\
\end{array}
\right) \label{eq:lagr-compact}
\end{align}
where%
\begin{subequations}
\begin{align}
&M_{B} =\left(
\begin{array}
[c]{cccc}%
-\square-4 M_{U,Q} \delta_{\pi R/2} & \partial_{y}+\eta\left(  y\right)  M_{U,Q} & 0 & \lambda_{t}%
\alpha^*\\
-\partial_{y}+\eta\left(  y\right)  M_{U,Q} & 1 & 0 & 0\\
0 & 0 & -\square+4 M_{Q,U} \delta_{\pi R/2}  & -\partial_{y}+\eta\left(  y\right)  M_{Q,U}\\
\lambda_{t}\alpha & 0 & \partial_{y}+\eta\left(  y\right)  M_{Q,U} &
1
\end{array}
\right)  \\
&M_{F}  =\left(
\begin{array}
[c]{cccc}%
\partial_{y}-\eta\left(  y\right)  M_{U,Q} & i\sigma^{\mu}\partial_{\mu} & 0 &
0\\
i\overline{\sigma}^{\mu}\partial_{\mu} & -\partial_{y}-\eta\left(  y\right)
M_{U,Q} & 0 & \lambda_{t}\alpha^* \\
\lambda_{t}\alpha & 0 & -\partial_{y}-\eta\left(  y\right)  M_{U,Q} & i\sigma^{\mu
}\partial_{\mu}\\
0 & 0 & i \overline{\sigma}^{\mu}\partial_{\mu} & \partial
_{y}-\eta\left(  y\right)  M_{U,Q}%
\end{array}
\right)  
\end{align} \label{eq:lagr-matrici}
\end{subequations}
\begin{equation*}
\begin{array}{lcr}
\alpha  = \frac{1}{2} \left[  \delta\left(  y\right)  +\delta\left(
y-\pi R\right)  \right]  v \; h^{\left( 0 \right)} \left( y=0 \right),
& \ \ & \delta_{\pi R/2}=\delta \left( y- \frac{\pi R}{2}\right)
\end{array}
\end{equation*}
with $h^{\left(0\right)}\left(  y=0\right)  $ the Higgs zero mode wave
function (\ref{eq:funzioned'ondahiggs}) at $y=0.$

Taking for the wave functions the form $\left(  \ref{eq:funzionid'onda}
\right)  $ (in this case one must consider also the wave functions
of $F_{U,Q},F_{Q,U}^{c}$) and imposing the proper boundary
conditions one can get
the equations for the masses of the fields $\varphi_{U,Q},\psi_{U,Q}%
,\varphi_{U,Q}^{c},\psi_{U,Q}^{c}.$

For the top quark $\psi_{U,Q}$ and the top squark $\varphi_{U,Q}$ (the
lowest modes) one gets%
\begin{equation}
m_{t}^{2} = \frac{\lambda_{t}^{2}v^{2}\left| h^{\left( 0 \right)}
    \left( 0 \right) \right|  ^{2}}{16}  \left( k_{U}^t \coth
    \frac{k_{U}^t \pi R}{2}-M_{U} \right)  
\left(k_{Q}^t \coth \frac{k_{Q}^t \pi R}{2}-M_{Q}
    \right)\label{eq:mtoplambdatop}
\end{equation}
\begin{align}
\left( k_{U}^{\tilde{t}} \coth \frac{k_{U}^{\tilde{t}} \pi
    R}{2}+M_{U} \right) \left( k_{Q}^{\tilde{t}} \coth
  \frac{k_{Q}^{\tilde{t}} \pi R}{2}-M_{Q} \right) = &
\frac{\lambda_{t}^{2} v^2 \left| h^{\left( 0 \right)} \left( 0 \right)
  \right|^2}{16} \nonumber  \\ & \left( m_{\tilde t}^2 +2 M_{Q} \left(k_Q^{\tilde t}
    \coth \frac{k_{Q}^{\tilde{t}} \pi R}{2}-M_{Q} \right) \right) 
\label{eq:mstoplambdatop}
\end{align}
where $k_{U,Q}^{t,\tilde{t}}=\sqrt{M_{U,Q}^{2}-m_{t,\tilde{t}}^{2}}.$ The wave function of
the top quark zero mode, normalized to $\int_{0}^{2 \pi R} \textrm{d}y \left|
\psi_{0}^{U,Q} \left(  y\right) \right|^2=1$, 
is obtained by solving equation (\ref{eq:equazionemodozeromateria}). For $y \in \left[ 0 , \pi R / 2 \right)$ we have
\begin{equation}
\psi_{0}^{U,Q}\left(  y\right)  =\frac{k_{U,Q}\cosh k_{U,Q}\left(  \frac{\pi
R}{2}-y\right)  -M_{U,Q}\sinh k_{U,Q}\left(  \frac{\pi R}{2}-y\right)  }%
{\sqrt{-m_t^2  \pi R+2 M_{U,Q} \left(- \cosh
k_{U,Q}\pi R + 1 \right) +\left(  k_{U,Q}+\frac{M^{2}}{k_{U,Q}}\right)  \sinh k_{U,Q}\pi R}%
}\label{eq:funzioned'ondatop}%
\end{equation}

The usual 4D top Yukawa coupling, defined as the coupling among
the zero
modes of the fields $h,\psi_{U},\psi_{Q},$ is given by%
\begin{equation}
y_{t}= \widehat{y}_t \eta_{0}^{h}\eta_{0}^{U}\eta_{0}^{Q}%
\label{eq:uaiti}
\end{equation}
where $\widehat{y}_t=\lambda_t \left( 2 \pi R \right)^{-3/2}$ and $\eta_{0}^i$ ($i=h,U,Q$) are the wave functions $\left(
\ref{eq:funzioned'ondahiggs}\right)  -\left(  \ref{eq:funzioned'ondatop}%
\right)  $ at $y=0$ normalized in such a way that
\begin{equation}
\int_{0}^{2 \pi R}\textrm{d}y \left| \eta^i_0\left( y \right) \right| ^2 = 2 \pi R
\end{equation}
Inserting (\ref{eq:uaiti}) into (\ref{eq:mtoplambdatop}) we obtain
the relation between the top quark mass $m_t$ and the 4D Yukawa
coupling $y_t$.

\section{Propagators}\label{sec:app-propag}

In order to obtain the mixed momentum--coordinate space propagators
for the components $\varphi$, $\psi$, $F$ of a hypermultiplet, we
start from the 5D Lagrangian without eliminating the auxiliary
fields. If $M \neq 0$ the relevant part of this Lagrangian is:
\begin{align}
{\cal L}= & \abs{\partial_{\mu} \varphi}^2+\abs{\partial_{\mu}
\varphi^c}^2+ \abs{F}^2+\abs{F^c}^2 + i \bar{\psi}
\bar{\sigma}^{\mu}\partial_{\mu}\psi + i \psi^c
\sigma^{\mu}\partial_{\mu}\bar{\psi}^c + \nonumber \\
&+ \left[ F \partial_y \varphi^c - F^c \partial_y \varphi + \hc \right]
+ M \eta(y) \left[ F \varphi^c + F^c \varphi + \hc
\right]  \nonumber \\
& + \left[\psi^c \partial_y \psi + \hc \right] - M \eta(y) \left[
\psi^c \psi + \hc \right] \nonumber \\
& -4 M \delta(y -\pi R/2) \left[ \abs{h}^2 -\abs{h^c}^2 \right]\label{eq:lagr-prop}
\end{align}
for the generic hypermultiplet of components $\varphi$, $\psi$, $F$
and their conjugates. The border term at $\pi R/2$ in the last line is
necessary to maintain 5D SUSY invariance under $\xi^{-+}$
transformations\footnote{Using the formulation in
terms of 4D N=1 superfields \cite{Arkani-Hamed:2001tb} one privileges only one of the two
local supersymmetries, here $\xi^{+-}$. This explains the apparent
asymmetry between $y=0$ and $y=\pi R/2$.}.
In this paper we need only the propagators for the
top--stop sector, so we can assume from now on that the parities
are those of the matter multiplets. Using the vectors defined in (\ref{eq:def-vect-lagr}),
$\mathcal L$ can be recast in a more compact form as in Appendix
\ref{sec:app-spectrum}:
\begin{equation}
{\cal L} = X^{\dagger} A X + Y^{\dagger} B Y + \bar{Z} C Z
\label{eq:lagr-prop-compact}
\end{equation}
where
\begin{align}
& \begin{array}{cc}
A=\left(\begin{array}{cc} - \square -4M \delta(y -\pi R/2) & \partial_y + M \eta(y)\\
-\partial_y + M \eta(y) & 1 \end{array}\right) &
B=\left(\begin{array}{cc} - \square  +4 M\delta(y -\pi R/2) & -\partial_y + M \eta(y) \\
\partial_y + M \eta(y) & 1
\end{array}\right) \end{array} \nonumber \\
& C= \left(\begin{array}{cc}  \partial_y - M \eta(y) & i \sigma^{\mu} \partial_{\mu}\\
i \bar{\sigma}^{\mu} \partial_{\mu} &  - \partial_y - M \eta(y)
\end{array}\right)
\end{align}
Note that the components of $X$ (or $Y$, $Z$) have the same
quantum numbers but different boundary conditions.

Let us focus, for example, on the propagator
\begin{equation}
G \left[(x-x')_{\mu};y,y' \right]=\langle \varphi (x_{\mu}',y')
\varphi^{\dagger} (x_{\mu},y) \rangle, \label{eq:def-prop}
\end{equation}
the others being analogous. In general all the correlation
functions will depend on both $y$ and $y'$\footnote{Or on $\bar y = (y+y')/2$ and $\Delta y=(y-y')$.}
because of the non conservation of the 5th component of momentum
in the segment $[0,\pi R /2 ]$. However, being interested in
calculating only loops formed using Yukawa interactions which are
localized at $y=0$, we can impose without any problem $y'=0$ from
the very beginning of the calculation, reducing the dependence of
the (\ref{eq:def-prop}) only to $(x-x')_{\mu}$ and $y$.

One can arrange propagators in matrices using the vectors
previously defined. In particular, defining
\begin{eqnarray}
{\cal G}(x-x';y) &=& \langle X(x,y) X^{\dagger}(x',0) \rangle
\nonumber \\ &=& \left(
\begin{array}{cc} \langle \varphi (x,y) \varphi^{\dagger} (x',0) \rangle & \langle
\varphi (x,y) F^c (x',0) \rangle \\ \langle F^{c \dagger} (x,y)
\varphi^{\dagger} (x',0) \rangle & \langle F^{c \dagger}  (x,y) F^c
(x',0) \rangle \end{array} \right)
\end{eqnarray}
the equations of motion for the scalar Green
functions are:
\begin{equation}
A {\cal G}(x-x';y) = i \delta^{(4)}(x-x') \delta (y) 
\label{eq:green-motion}
\end{equation}

Multiplying the 1st row of $A$ by the 1st column of $\mathcal G$
we get a system of 2 differential equations, which after passing
to euclidian 4-momentum, assumes the form:
\begin{equation}
\left\{\begin{array}{l}
-k_4^2 g(y) -4M \delta(y -\pi R/2) + \left(\partial_y + M \eta(y) \right)f(y) =\delta(y) \nonumber\\
(-\partial_y + M \eta(y))g(y) + f(y)=0
\end{array} \right.
\end{equation}
where $g(y)=\langle \varphi (y) \varphi^{\dagger}(0) \rangle$ and
$f(y)=\langle F^{c \dagger} (y) \varphi^{\dagger}(0) \rangle$. These coupled
equations must be solved imposing the $(+,-)$ and the $(-,+)$ boundary conditions
in $y$ on $g$ and $f$ respectively, using the same techniques of Appendix
\ref{sec:app-spectrum}. One finally gets the $\langle \varphi
\varphi^{\dagger} \rangle$ propagator $G_{\varphi} \left( k_4,y;M \right)$:
\begin{equation}
G_{\varphi} \left( k_4,y;M \right) = \langle \varphi \varphi^{\dagger} \rangle (y) = \frac{\sinh
\left[k\left(\frac{\pi R}{2}-y \right)\right]}{2 \left[ k \cosh
\left(\frac{k \pi R}{2} \right)+M \sinh \left(\frac{k \pi R}{2}
\right)\right]} \label{eq:prop-M0-1}
\end{equation}
where $k=\sqrt{k_4^2+M^2}$. Analogously the $\langle F F^{\dagger}
\rangle $ $(+,-)$ and $\langle \psi \psi^{\dagger} \rangle $
$(+,+)$ propagators are:
\begin{align}
& G_{F} \left( k_4,y;M \right) =  \frac{k_4^2}{2}\frac{ \sinh
\left[k\left(\frac{\pi R}{2}-y \right)\right]}{ \left[ k \cosh
\left(\frac{k \pi R}{2} \right)-M \sinh \left(\frac{k \pi R}{2}
\right)\right]} \left(1- \frac{2 M}{k_4^2}\left(k \coth \left(\frac{k \pi
  R}{2} \right) -M\right)\right)
\nonumber\\
& G_{\psi} \left( k_4,y;M \right) =  \frac{\sla{k}_4}{2
k_4^2} \frac{  k \cosh \left[k\left(\frac{\pi R}{2}-y
\right)\right] +M \sinh \left[k\left(\frac{\pi R}{2}-y
\right)\right]}{\sinh \left(\frac{k \pi R}{2}\right)}
\label{eq:prop-M0-2}
\end{align}
where $\sla{k}_4=\sigma \cdot k_4$.

In the limit $M \rightarrow 0$ these propagators become:
\begin{align}
& G_{\varphi} (k_4,y;M=0)= \frac{1}{2 k_4} \frac{\sinh
\left[k_4\left(\frac{\pi R}{2}-y \right)\right]}{\cosh
\left(\frac{k_4 \pi R}{2}
\right)} \nonumber\\
& G_{F} (k_4,y;M=0) = \frac{k_4}{2} \frac{ \sinh
\left[k_4\left(\frac{\pi R}{2}-y \right)\right]}{\cosh
\left(\frac{k_4 \pi R}{2}
\right)} \nonumber\\
& G_{\psi} (k_4,y;M=0) = \frac{\sla{k}_4}{2 k_4} \frac{  \cosh
\left[k_4\left(\frac{\pi R}{2}-y \right)\right] }{\sinh
\left(\frac{k_4 \pi R}{2}\right)} \label{eq:prop-M0-v0}
\end{align}

\section*{Acknowledgments}
We thank Roberto Contino, Paolo Creminelli, Lawrence Hall, Federico
Minneci, Takemichi
Okui, Steven Oliver and Riccardo
Rattazzi for useful discussions. This work has been partially supported by the EC under
TMR contract HPRN-CT-2000-00148.


\end{document}